\documentclass[12pt]{article}
\usepackage{amssymb,amsmath,epsfig}

\begin{document}

\title{\bf Dynamical Stability of Collapsing Stars in Einstein Gauss-Bonnet Gravity}

\author{ G. Abbas
\thanks{ghulamabbas@ciitsahiwal.edu.pk}, S. Sarwar
\thanks{shahzadch@ciitsahiwal.edu.pk}
\\Department of Mathematics, COMSATS Institute\\
of Information Technology, Sahiwal-57000, Pakistan.}
\date{}
\maketitle
\begin{abstract}
This paper is devoted to study the stability/instability of the
expansionfree self gravitating source in the framework of Einstein
Gauss-Bonnet gravity. The source has been taken as Tolman-Bondi
model which is homogenous in nature. The field equations as
dynamical equations have been evaluated in Gauss-Bonnet gravity in
five dimensions. The junction conditions as well as cavity
evaluations equations have been explored in detail. The perturbation
scheme of first order has been applied to dynamical as Einstein
Gauss-Bonnet field equations. The concept of Newtonian as well post
Newtonian approximation have been used to derive general dynamical
stability equations. In general this equation represents the
stability of the gravitating source. Some particular values of
system parameters have been chosen to prove the concept of stability
graphically. It has been mentioned that other than chosen the
particular values of the parameters the stability of the system will
be disturbed , hence it would leads to instability.

\end{abstract}

\

\section{Introduction}
The dynamical instability of the astrophysical objects is the
subject of interest in classical physical as well as in general
theory of relativity (GR). The motivation of this problem become
important when static stellar models are stable against the
fluctuations produced by the self gravitational attraction of the
massive stars. It is most relevant to structure formation during the
different phases of the gravitationally collapsing objects. In the
relativistic astrophysics the dynamical stability of the stars was
studied by the Chandrasekhar \cite{1} in 1964 , since then a
renowned interest has grown in this research area. Herrera et
al.\cite{2,3} have extended the pioneers work for non-adiabatic,
anisotropic and viscous fluids. All these investigations imply that
adiabatic index ${\Gamma}_1$ define the range of instability, for
example for the Newtonian perfect fluid such range is
${\Gamma}_1<4/3$. Friedman \cite{4} discussed the dynamical
instability of neutral fluid sphere in the Newtonian as well as in
the relativistic physics and showed that anisotropy enhances the
stability if anisotropy is positive throughout matter distribution.
Herrera et al. \cite{5} have studied the dynamical instability of
the expansion free fluids using perturbation scheme.

Different physical properties of the fluid plays important role in
dynamical evaluation of the self-gravitating systems.  According to
Herrera et al. \cite{6,7} dissipation terms in the fluid would
increase the instability of the collapsing objects. Chan and his
collaborators \cite{8}-\cite{11} have showed that anisotropy and
radiation would affect the instability range at Newtonian nd
post-Newtonian approximation. Sharif and Azam \cite{11a}-\cite{14}
have studied the effects of electromagnetic field on the dynamical
stability of the collapsing dissipative and non dissipative fluids
in spherical, cylindrical and plane symmetric geometries. This work
has been further extended by Sharif and his collaborators
\cite{15}-\cite{24} in higher order theories of gravity, like $f(R)$
and $f(T)$ and $f(R,T)$, in these papers the possible forms of the
fluid with electromagnetic field have been discussed in detail.

Till now many quantum theories of gravity have been proposed to
investigate the natural phenomenon occurring in astronomy and
astrophysics. Among these theories, superstring theory is the most
strong candidate which has been extensively investigated for the
spacetimes with more than four dimensions. In this theory the
effects of extra dimensions becomes more prominent when curvature
radius of the central high density regions during gravitational
collapse becomes comparable with the curvature radius of the extra
dimensions.From this  point of view high density regions can be
modeled in a sophisticated way in a  theory which deals the extra
dimensions. The braneworld universe model which is attractive
proposal for the new picture of the universe is based on the
superstring theory \cite{25}-\cite{30}. The geometrical
interpretation of the braneworld model revels the fact that we are
living on a four dimensional timelike hypersurface which is embedded
in more than four dimensional manifold. This suggest that effects of
superstring on the formation of back hole during the relativistic
gravitational collapse of a star should be investigated explicitly.

The current experiments performed for the tests of inverse square
law do not exclude the possibility of the extra dimension even as
large as a tenth of millimeter. As observed range of the
gravitational force is directly dependent on the size of objects so
it is interesting to consider some physical phenomena in the extra
dimensions.On the basis of these facts it becomes important to study
the general theory of relativity in more than four dimensions. In
this regards a class of exact solutions to the Einstein field
equations have been determined in the recent years
\cite{31}-\cite{35}. These solutions play a significant role in
studying the gravitational collapse evolution of the universe.
Recently \cite{36}-\cite{43} there has been growing interest to
study the higher order gravity , which are involves higher order
derivatives of curvature terms. One of the most studied extensively
higher order gravity theory is the Gauss-Bonnet gravity. This theory
is the simplest generalization of general theory of relativity and
special case of Lovelock Gravity theory. The Lagrangian of this
theory contains just three terms as compared to Lagrangian of
Loelock gravity Theory.

The Gauss-Bonnet gravity theory is used to discuss the nontrivial
dynamical systems in the dimensions greater or equal to 5. This
theory naturally appears in the low energy effective action of the
heterotic string theory. Boulware and Deser \cite{44} formulated the
black hole (BH) solutions in N dimensional gravitational theory with
four dimensional Gauss-Bonnet term. These are generalization of N
dimensional solutions investigated by Tangherili \cite{45}, Merys
and Perry \cite{46}. The spherically symmetric BH solutions and
their physical properties have been studied in detail by Wheeller
\cite{47}. The structure of topologically nontrivial BHs has been
presented by Cai \cite{48}. Kobayashi \cite{49} and Maeda \cite{50}
have explored the effects of Gauss-Bonnet term on the structure of
Vaidya BH. All these studies show that appearance of the
Gauss-Bonnet term in the field Equations would effect the occurrence
of BH and Naked singularity during the gravitational collapse. In a
recent paper \cite{52}, Jhinag and Ghosh have consider the $5D$
action with the Gauss-Bonnet terms in Tolman-Bondi model and give an
exact model of the gravitational collapse of a inhomogeneous dust.
Motivated by these studies, we have discussed the stability of the
gravitationally collapsing spheres in Einstein Gauss-Bonnet gravity.
This paper is organized as follow: In section \textbf{{2}} the
Einstein Gauss-Bonnet field equations and dynamical equations have
been presented. The perturbation scheme of first order on the field
equations as well as on dynamical equations have been presented in
section \textbf{3}. Section\textbf{ 4} deals with the Newtonian and
post Newtonian approximation and derivation of the stability
equation, which is main result of the paper. We  summaries the
results of the paper in the last section.

\section{Interior Matter Distribution and Einstein Gauss-Bonnet Field Equations }

We begin with the following 5D action:
\begin{equation}\label{1}
S=\int d^{5}x\sqrt{-g}\left[ \frac{1}{2k_{5}^{2}}\left( R+\alpha
L_{GB}\right) \right] +S_{matter}
\end{equation}%
where $R$ ia a $5D$ Ricci scalar and $k^2_{5}={8\pi G_{5}}$ is $5D$
gravitational constant. The Gauss-Bonnet Lagrangian is of the form
\begin{equation}\label{2}
L_{GB}=R^{2}-4R_{ab}R^{ab}+R_{abcd}R^{abcd}
\end{equation}
where $\alpha$ is the coupling constant of the Gauss-Bonnet terms.
This type of action is derived in the low-energy limit of heterotic
superstring theory. In that case, $\alpha $ is regarded as the
inverse string tension and positive definite and we consider only
the case with $\alpha \geq 0$ in this paper. In the $4D$ space-time,
the Gauss-Bonnet terms do not contribute to the Einstein field
equations. The action (\ref{1}) leads to the following set of field
equations
\begin{equation}\label{3}
{G}_{ab}=G_{ab}+\alpha H_{ab}=T_{ab},
\end{equation}
where
\begin{equation}\label{4}
G_{ab}=R_{ab}-\frac{1}{2}g_{ab}R
\end{equation}
is the Einstein tensor and
\begin{equation}\label{5}
H_{ab}=2\left[ RR_{ab}-2R_{a\alpha }R_{b}^{\alpha }-2R^{\alpha \beta
}R_{a\alpha b\beta }+R_{a}^{\alpha \beta \gamma }R_{b\alpha \beta \gamma }%
\right] -\frac{1}{2}g_{ab}L_{GB},
\end{equation}
is the Lanczos tensor.

A spacelike 4D hypersurface $\Sigma^{(e)}$ is taken such that it
divides a 5D spacetime into two 5D manifolds, $M^-$ and $M^+$,
respectively. The 5D TB spacetime is taken as an interior manifold
$M^-$  which represents an interior of a collapsing inhomogeneous
and anisotropic sphere is given by \cite{52}
\begin{equation}\label{6}
ds_{-}^2=-dt^2+A^2dr^2+R^2(d\theta^2+\sin^2{\theta}d\phi^2
+\sin^2{\theta}\sin^2{\phi}d\psi^2),
\end{equation}
where $A$ and $R$ are functions of $t$ and $r$. The energy-momentum
tensor $T_{\alpha \beta }^{-}$ for anisotropic fluid has the form
\begin{equation}\label{7}
T_{\alpha \beta }^{-}=(\mu +P_{\perp })V_{\alpha }V_{\beta
}+P_{\perp }g_{\alpha \beta }+(P_{r}-P_{\perp })\chi _{\alpha }\chi
_{\beta },
\end{equation}
where $\mu $ is the energy density, $P_{r}$ the radial pressure,
$P_{\perp }$the tangential pressure, $V^{\alpha }$ the four velocity
of the fluid and $ \chi _{\alpha }$ a unit four vector along the
radial direction. These
quantities satisfy,%
\begin{equation}
V^{\alpha }V_{\alpha }=-1\ \ ,\ \ \ \ \ \chi ^{\alpha }\chi _{\alpha
}=1\ \ ,\ \ \ \ \ \chi ^{\alpha }V_{\alpha }=0  \label{N8}
\end{equation}%
The expansion scalar $\Theta $ for the fluid is given by
\begin{equation}\label{8}
\Theta =V_{\ ;\ \alpha ,}^{\alpha }.
\end{equation}%
Since we assumed the metric (6) comoving, then
\begin{equation}\label{9}
V^{\alpha }=A^{-1}\delta _{0}^{\alpha }\ ,\ \ \ \ \ \chi ^{\alpha
}=B^{-1}\delta _{1}^{\alpha }\
\end{equation}%
and for the expansion scalar, we get
\begin{equation}\label{10}
\Theta =\frac{\dot{A}}{A}+\frac{3\dot{R}}{R}.
\end{equation}
Hence, Einstein Gauss-Bonnet field equations take the form%
\begin{eqnarray}\nonumber
k^2_{5}\mu &&=\frac{12\left( R^{\prime 2}-A^{2}\left(
1+\dot{R}^{2}\right) \right) }{R^{3}A^{5}}\left[ R^{\prime
}A^{\prime }+A^{2}\dot{R}\dot{A}-AR^{\prime \prime }\right] \alpha
\\\label{11} &&\ \ \ \ \ -\frac{3}{A^{3}R^{2}}\left[ A^{3}\left(
1+\dot{R}^{2}\right) +A^{2}R\dot{R}\dot{A}+RR^{\prime }A^{\prime
}-A(RR^{\prime \prime }+R^{\prime 2})\right]\\\label{12a}
k^2_{5}p_{r} &&=-12\alpha \left( \frac{1}{R^{3}}-\frac{R^{^{\prime }2}}{A^{2}R^{3}}%
+\frac{\dot{R}^{2}}{R^{3}}\right) \ddot{R}+3\frac{R^{^{\prime }2}}{A^{2}R^{2}%
}  -3\Big(\frac{1+\dot{R}^{2}+R\ddot{R}}{R^{2}}\Big)\\\nonumber
k^2_{5}p_{\perp } &&=\frac{4\alpha }{A^{4}R^{2}}\Big[ -2A\left(
A^{^{\prime
}}R^{^{\prime }}+A^{2}\dot{A}\dot{R}-AR^{^{\prime \prime }}\right) \ddot{R}%
 +A\left( R^{^{\prime
}2}-A^{2}\left( 1+\dot{R}^{2}\right) \right)
\ddot{A}\\\nonumber&&+2\Big( \dot{A}R^{^{\prime
}}-A\dot{R}^{^{\prime }}\Big] -\frac{1}{A^{3}R^{2}}\Big[ A^{3}\Big(
1+\dot{R}^{2}+2R\ddot{R}\Big) +A^{2}R\left(
2\dot{R}\dot{A}+R\ddot{A}\right)\\&&+2RR^{^{\prime }}A^{^{\prime
}}-2A\left( RR^{^{\prime \prime }}+R^{^{\prime }2}\right)\Big]
\label{13}\\
 &&\frac{12\alpha }{A^{5}R^{3}}\left( \dot{A}R^{^{\prime
}}-A\dot{R}^{^{\prime
}}\right) \left( A^{2}\left( 1+\dot{R}^{2}\right) -R^{^{\prime }2}\right) -3%
\frac{A\dot{R}^{^{\prime }}-\dot{A}R^{^{\prime }}}{A^{3}R}=0
\label{14}
\end{eqnarray}
The mass function $m(t,r)$ analogous to Misner-Sharp mass in $n$
manifold without ${\Lambda}$ is given by \cite{50}
\begin{equation}\label{15}
m(t,r)=\frac{(n-2)}{2k_{n}^{2}}{V^k}_{n-2}\left[ R^{n-3}\left(
k-g^{ab}R,_{a}R,_{b}\right) +(n-3)(n-4)\alpha \left(
k-g^{ab}R,_{a}R,_{b}\right) ^{2} \right],
\end{equation}
where a comma denotes partial differentiation and ${V^k}_{n-2}$ is
the surface of $(n-2)$ dimensional unit space. For $k=1$,
${V^1}_{n-2}=\frac{2{\pi}^{(n-1)/2}}{\Gamma((n-1)/2)}$, using this
relation with $n=5$ and Eq.(\ref{6}), the mass function (\ref{15})
reduces to
\begin{equation}\label{16}
m(r,t)=\frac{3}{2}\left[ R^{2}\left( 1-\frac{R^{^{\prime }2}}{A^{2}%
}+\dot{R}^{2}\right) +2\alpha \left( 1-\frac{R^{^{\prime }2}}{A^{2}}+\dot{R}%
^{2}\right) ^{2}\right]
\end{equation}
The nontrivial components of the Binachi identities, $T_{;\beta }^{\
-\alpha \beta }=0$, from Eqs.(\ref{6}) and (\ref{7}), yield
\begin{equation}
\left[ \dot{\mu}+\left( \mu +P_{r}\right) \frac{\dot{A}}{A}+3\left(
\mu +P_{\perp }\right) \frac{\dot{R}}{R}\right] =0 , \label{17}
\end{equation}%
and%
\begin{equation}
T_{;\beta }^{\ -\alpha \beta }\chi _{\alpha }=\frac{1}{A}\left[
P_{r}^{^{\prime }}+3\left( P_{r}-P_{\perp }\right) \frac{R^{^{\prime }}}{R}%
\right] =0  \label{18}
\end{equation}

Using field equations and Eq.(\ref{16}), we may write

\begin{equation}
m^{\prime }=\frac{2}{3}k^2_{5}\mu R^{\prime }R^{3}  \label{N16a}
\end{equation}
In the exterior region to $\Sigma^{(e)}$, we consider Einstein
Gauss-Bonnet Schwarzschild solution which is given by \cite{54}

\begin{equation}\label{c1}
ds_{+}^2=-F(\rho)d{\nu}^2-2d\nu
d\rho+\rho^2(d\theta^2+\sin^2{\theta}d\phi^2
+\sin^2{\theta}\sin^2{\phi}d\psi^2),
\end{equation}
where
$F(\rho)=1+\frac{{\rho}^2}{4\alpha}-\frac{{\rho}^2}{4\alpha}\sqrt{1+\frac{16\alpha
M}{\pi {\rho}^4}}$

The smooth matching of the $5D$ anisotropic fluid sphere (\ref{6})
to GB Schwarzschild BH solution (\ref{c1}), across the interface at
$r = {r_{\Sigma}}^{(e)}$ = constant, demands the continuity of the
line elements and extrinsic curvature components (i.e., Darmois
matching conditions), implying
\begin{eqnarray}\label{c2}
dt \overset{\Sigma^{(e)}}{=}\sqrt{F(\rho)}d\nu,\\
R \overset{\Sigma^{(e)}}{=}\rho, \\\label{cm}
m(r,t)\overset{\Sigma^{(e)}}{=}M,
\end{eqnarray}
\begin{eqnarray}\nonumber
 &&-12\alpha \left( \frac{1}{R^{3}}-\frac{R^{^{\prime }2}}{A^{2}R^{3}}%
+\frac{\dot{R}^{2}}{R^{3}}\right) \ddot{R}+3\frac{R^{^{\prime }2}}{A^{2}R^{2}%
}  -3\Big(\frac{1+\dot{R}^{2}+R\ddot{R}}{R^{2}}\Big)\\
 &&\overset{\Sigma^{(e)}}{=}\frac{12\alpha }{A^{5}R^{3}}\left( \dot{A}R^{^{\prime
}}-A\dot{R}^{^{\prime
}}\right) \left( A^{2}\left( 1+\dot{R}^{2}\right) -R^{^{\prime }2}\right) -3%
\frac{A\dot{R}^{^{\prime }}-\dot{A}R^{^{\prime }}}{A^{3}R}
\label{c3}
\end{eqnarray}
Comparing Eq.(\ref{c3}) with (\ref{12a}) and (\ref{14}) (for detail
see \cite{12}), we get
\begin{equation}\label{c4}
p_r\overset{\Sigma^{(e)}}{=}0.
\end{equation}
Hence, the matching of the interior inhomogeneous anisotropic fluid
sphere (\ref{6}) with the exterior vacuum Einstein Gauss-Bonnet
spactime (\ref{c1}) produces Eqs.(\ref{6}) and (\ref{cm}). These are
the necessary and sufficient conditions for the smooth matching of
interior and exterior regions of a star on boundary surface
${\Sigma^{(e)}}$.

It is well known that the expansionfree models present an internal
vacuum cavity. The boundary surface between the external cavity and
interior the fluid is labeled by ${\Sigma^{(i)}}$ then the smooth
matching of the Minkowski spacetime within the cavity to the fluid
distribution over ${\Sigma^{(i)}}$, yields

\begin{eqnarray}
m(r,t)\overset{\Sigma^{(i)}}{=}0.\\
p_r\overset{\Sigma^{(i)}}{=}0.
\end{eqnarray}
The physical applications of expansionfree models are wide in
astrophysics and astronomy. For example, it may help to explore the
structure of voids on cosmological scales\cite{55}. By definition
Voids are the sponge like structures and occupying 40-50 percent of
the entire universe. There are commonly two types of the voids:
mini-voids \cite{56} and macro-voids\cite{57} On the basis of
Observational data analysis the voids are neither empty nor
spherical. For the sake of further exploration about voids they are
considered as vacuum spherical cavities around the fluid
distribution.

\section{The Perturbation Scheme }

In this section, we introduce the perturbation scheme, for this
purpose it is assumed that initially fluid is in static equilibrium
implying that the fluid is described by only such quantities which
have only radial dependence. Such quantities are denoted by a
subscript zero. We further assume as usual, that the metric
functions $A(t,r)$ and $R(t,r)$ have the same time dependence in
their perturbations. Therefore, we consider the metric and material
functions in the following form
\begin{eqnarray}
A(t,r)&=&A_{0}(r)+\epsilon T(t)a(r), \label{N17}\\
R(t,r)&=&R_{0}(r)+\epsilon T(t)c(r),  \label{N19}\\
 \mu (t,r)&=&\mu
_{0}(r)+\epsilon \bar{\mu}(t,r),  \label{N20}\\
P_{r}(t,r)&=&P_{r0}(r)+\epsilon \bar{P}_{r}(t,r),  \label{N21}\\
P_{\perp }(t,r)&=&P_{\perp 0}(r)+\epsilon \bar{P}_{\perp }(t,r),
\label{N22}\\
 m(t,r)&=&m_{0}(r)+\epsilon \bar{m}(t,r),  \label{N23}\\
\Theta (t,r)&=&\epsilon \bar{\Theta}(t,r), \label{N24}
\end{eqnarray}
where $0<\epsilon \ll 1$ and we choose the Schwarzschild coordinates
with $R_{0}(r)=r$. Using Eqs.(\ref{N17})-(\ref{N22}), we have from
Eqs.(\ref{11})-(\ref{14}) the following static configuration
\begin{eqnarray}
k\mu _{0}&&=\frac{3}{r^{3}A_{0}^{4}}\left[ 4\alpha \left( \frac{%
A_{0}^{^{\prime }}}{A_{0}}-A_{0}\right) -rA_{0}^{2}\left( A_{0}+\frac{%
A_{0}^{^{\prime }}}{A_{0}}r-1\right) \right], \label{N25}\\
kP_{r0}&&=3\left[ \frac{1}{r^{2}A_{0}^{2}}-\frac{1}{r^{2}}-1\right],
\label{N26}\\
kP_{\perp 0}&&=\frac{-1}{A_{0}^{2}r^{2}}\left[ A_{0}^{2}+2r\frac{%
A_{0}^{^{\prime }}}{A_{0}}-2\right].  \label{N27}
\end{eqnarray}
Also from Eqs.(\ref{11})-(\ref{14}), we obtain the following form of
the perturbed field equations
\begin{eqnarray}\nonumber
k\bar{\mu} &&=\frac{3T}{r^{2}A_{0}^{3}}\Big[ 4\alpha \Big(
\frac{a^{\prime
}}{rA_{0}^{2}}+\frac{3A_{0}^{\prime }c^{\prime }}{rA_{0}^{2}}-\frac{%
c^{\prime \prime }}{rA_{0}}-\frac{a^{\prime
}}{A_{0}r}-\frac{c^{\prime }A_{0}^{\prime }}{rA_{0}} \\\nonumber
&&+ \frac{c^{\prime \prime }}{r}+\frac{3aA_{0}^{\prime }}{rA_{0}^{2}}-%
\frac{5aA_{0}^{\prime }}{rA_{0}^{2}}-\frac{cA_{0}^{\prime }}{r^{2}A_{0}^{2}}+%
\frac{cA_{0}^{\prime }}{r^{2}A_{0}}\Big)   \\\nonumber &&-\Big(
-3aA_{0}+2c-3ar\frac{A_{0}^{^{\prime }}}{A_{0}}+A_{0}^{\prime
}c+ra^{\prime }\\\label{N28}
 && +A_{0}^{\prime }rc^{\prime
}-A_{0}rc^{\prime \prime }-A_{0}+2a-2A_{0}c^{\prime }\Big) \Big]
-\frac{2Tc}{r}k\mu _{0}\\\nonumber
k\bar{P}_{r}&&=\frac{3\ddot{T}c}{r}\left[ 1-4\alpha \left( 1-\frac{1}{%
r^{2}A_{0}^{2}}\right) \right] -\frac{6T}{r^{2}}\left( \frac{a}{A_{0}^{3}}-%
\frac{c^{\prime }}{A_{0}^{2}}+cr\right) -\frac{2Tc}{r}kP_{r0},\\
\label{N29}\\\nonumber
k^2_5\bar{P}_{\perp }
&=&\frac{\ddot{T}}{A_{0}^{3}r^{2}}\left[ 4\alpha \left( a\left(
1-A_{0}^{2}\right) -2A_{0}^{^{\prime }}c\right) -A_{0}rc\left(
A_{0}^{2}+r\right) \right] \\\nonumber &&+\frac{8\alpha
\dot{T}}{A_{0}^{3}r^{2}}\left( \frac{a}{A_{0}}-c^{^{\prime
}}\right) +\frac{T}{A_{0}^{2}r^{2}}\left[ 2rc^{^{\prime }}\left( \frac{%
A_{0}^{^{\prime }}}{A_{0}}\right) +2a\left( \frac{A_{0}^{^{\prime }}}{A_{0}}%
\right) -2rc^{^{\prime \prime }}\right.   \\\label{N30} &&\left.
+2r\left( \frac{a}{A_{0}}\right) ^{^{\prime }}-4c^{^{\prime
}}r-5\left( \frac{a}{A_{0}}\right) -4ar\left( \frac{A_{0}^{^{\prime }}}{A_{0}%
}\right) \right] -\frac{2Tc}{r}kP_{\perp 0},  \\
&&\frac{12\alpha \dot{T}}{A_{0}^{5}r^{3}}\left(
A_{0}^{2}a-A_{0}^{3}c-a-A_{0}c\right)
-\frac{3\dot{T}}{A_{0}^{3}r}\left( A_{0}c^{\prime }-a\right) =0.
\label{N31}
\end{eqnarray}
For the expansion given in Eq.(\ref{10}), we have
\begin{equation}
\bar{\Theta}=\dot{T}\left( \frac{a}{A_{0}}+\frac{3c}{R_{0}}\right).
\label{N32}
\end{equation}%
The Binachi identities Eqs.(\ref{17}) and (\ref{18}) with
(\ref{N17})-(\ref{N22}), yield the static configuration
\begin{equation}
P_{r0}^{\prime }+\frac{3}{r}\left( P_{r0}-P_{\perp 0}\right) =0
\label{N33}
\end{equation}%
and for the perturbed configuration
\begin{eqnarray}
\frac{1}{A_{0}}\left[ \bar{P}_{r}^{\prime }+\frac{3}{r}\left( \bar{P}_{r}-%
\bar{P}_{\perp }\right) +3\left( P_{r0}-P_{\perp 0}\right) T\left( \frac{c}{r%
}\right) ^{\prime }\right] =0 , \label{N34}\\
\bar{\mu}=-\left[ \left( \mu _{0}+P_{r0}\right) \frac{a}{A_{0}}+\frac{3c}{r}%
\left( \mu _{0}+P_{\perp 0}\right) \right] T . \label{N35}
\end{eqnarray}%
The total energy inside $\Sigma^{(e)\text{ }}$ up to a radius $r$ given by Eq.(\ref%
{16}) with Eqs.(\ref{N17}),(\ref{N19}) and (\ref{N23}) becomes
\begin{eqnarray}
m_{0}&&=\frac{3}{2}\left[ \left( 1-\frac{1}{A_{0}^{2}}\right) \left(
r^{2}+2\alpha \left( A_{0}^{2}-1\right) \right) \right],
\label{N36}\\ \bar{m}&&=\frac{3T}{A_{0}^{2}}\left[ \left(
A_{0}^{2}cr-c-c^{^{\prime }}r^{2}+r^{2}\frac{a}{A_{0}}\right)
-\frac{\alpha }{A_{0}^{2}}\left( A_{0}^{2}-1\right) \left(
c^{^{\prime }}-\frac{a}{A_{0}}\right) \right]. \label{N37}
\end{eqnarray}
From the matching condition Eq.(\ref{c4}), we have
\begin{equation}
P_{r0}\overset{\Sigma^{(e)}}{=}0,\ \ \ \
\bar{P}_{r}\overset{\Sigma^{(e)}}{=}0,  \label{N38}
\end{equation}%
For $c\neq 0$, which is the case that we want to study , with (\ref{N29}),(%
\ref{N31}) and (\ref{N38}) we obtain
\begin{equation}
\ddot{T}\ \beta -\gamma T=0,  \label{N39}
\end{equation}
where
\begin{equation*}
\beta =1-4\alpha \left( 1-\frac{1}{r^{2}A_{0}^{2}}\right) ,\ \ \ \ \
\ \
\gamma =\frac{2}{rc}\left( \frac{a}{A_{0}^{3}}-\frac{c^{\prime }}{A_{0}^{2}}%
+rc\right)
\end{equation*}

The general solution of Eq.(\ref{N39}) is actually the linear
combination of two solutions one of these corresponding to stable
(oscillating) system while other corresponds to unstable
(non-oscillating) ones. As in the present case, we are interested to
establish the range of instability, so we restrict our attention to
the non oscillating ones, i.e., we assume that $a(r)$ and $c(r)$
attain such values on $r_{\Sigma ^{(e)}}$ that $\psi_{\Sigma ^{(e)}}
=\Big(\frac{\beta }{\gamma }\Big) _{\Sigma ^{(e)}}>0.$ Then
\begin{equation}
T=\exp (-\sqrt{\psi _{\Sigma ^{(e)}}}t)  \label{N40}
\end{equation}
representing collapsing sphere as areal radius becomes decreasing
function of time.

The dynamical instability of collapsing fluids can be well discussed
in term of adiabatic index $\Gamma _{1}$. We relate$\bar{P}_{r}$ and
$\bar{\mu}$ for the static spherically symmetric configuration as
follows
\begin{equation}
\bar{P}_{r}=\Gamma _{1}\frac{P_{r0}}{\mu _{0}+P_{r0}}\bar{\mu}.
\label{N41}
\end{equation}
We consider it constant throughout the fluid distribution or at
least, throughout the region that we want to study.

\section{Newtonian and Post Newtonian Terms and Dynamical Stability}
This section deals to identify the Newtonian (N), post Newtonian
(pN) and post post Newtonian (ppN) regimes. For this purpose we
convert the relativistic units into c.g.s. units and expands all the
terms in dynamical equations upto the $C^{-4}$ ($C$ being speed of
light). In this analysis for the different regimes following
approximation will be applicable
\begin{itemize}
\item N order: terms of order $C^0$;
\item  pN order: terms of order $C^{-2}$;
\item ppN order: terms of order $C^{-4}$.
\end{itemize}
These terms are analyzed for the stability conditions appearing in
the dynamical equation in the N approximations while pN and ppN are
neglected. Thus, for N approximation, we assume
\begin{equation}
\mu _{0}\gg P_{r0},\ \ \ \ \ \mu _{0}\gg P_{\perp 0}  \label{N42}
\end{equation}
For the metric coefficient expanded up to pN approximation, we take
\begin{equation}
A_{0}=1+\frac{Gm_{0}}{C^{2}r},  \label{N43}
\end{equation}
where $G$ is the gravitational constant and $C$ is the speed of
light. With the help of equations obtained in previous sections, we
can formulate the dynamical equation with expansion-free condition
which is aim of our study. The key equation for the dynamical
equation is Eq.(\ref{N34}).

The expansion-free condition $\Theta =0$ implies from (\ref{N32})
\begin{equation}
\frac{a}{A_{0}}=-3\frac{c}{r},  \label{N44}
\end{equation}
with (\ref{N44}), we have for (\ref{N35}) that

\begin{equation}
\bar{\mu}=3(P_{r0}-P_{\perp 0})T\frac{c}{r}.  \label{N45}
\end{equation}
This equation explains how perturbed energy density of the system
originates from the static background anisotropy.

Also, with (\ref{N41}) and (\ref{N45}) we have

\begin{equation}
\bar{P}_{r}=3\Gamma _{1}\frac{P_{r0}}{\mu _{0}+P_{r0}}(P_{r0}-P_{\perp })T%
\frac{c}{r}  \label{N46}
\end{equation}

From equations (\ref{N25})and (\ref{N36}), we have

\begin{equation}
\frac{A_{0}^{^{\prime
}}}{A_{0}}=\frac{(r+m_{0})[(r+m_{0})^{3}k^2_5\mu _{0}+12\alpha
]}{12\alpha r-3r(r+m_{0})}  \label{N47}
\end{equation}

Next, we develop dynamical equation by substituting Eq.(\ref{N30})
along with Eqs. (\ref{N44}),(\ref{N43}),(\ref{N39}) and (\ref{N47}) in Eq. (\ref%
{N34}) and using the radial functions $a(r)=a_{0}r,\ c(r)=c_{0}r$,
where $a_0$ and $c_0$ are  constants. After a tedious algebra ( a
detail procedure can be  followed in \cite{6}), we obtain the
dynamical equation at pN order ( with $c=G=1$)

\begin{eqnarray}\nonumber
&&\Big(12\alpha r-3r(r+m_{0})\Big)\Big[3\psi
(r+m_{0})^2\Big(12\alpha c_{0}m_{0}(m_{0}+2r)-96\alpha
^{2}c_{0}r^{3}(r+m_{0})  \\\nonumber
&&-c_{0}\Big((r+m_{0})^{2}+r^{3}\Big)\Big)+{8\alpha r^{8}\sqrt{%
\psi }c_{0}}+3r^{3}c_{0}{(r+m_{0})}\Big(4r-15\Big)
+6r^{3}(r+m_{0})^{3}c_{0}k^2_5P_{\perp 0} \\\nonumber
&&+3r^{3}(r+m_{0})^{3}k^2_5(P_{r0}-P_{\perp
0})\Big]+{(r+m_{0})}\Big[216\alpha
r^{3}c_{0}(1-2r)(r+m_{0})^{2}-72\alpha r^{4}c_{0}\Big] \\\nonumber
 &&=\Big[
24r^3\alpha \psi c_{0}{(r+m_{0})}+{6r^4c_{0}}
+18c_{0}r^3(2r-1){(r+m_{0})}\Big]k^2_5\lambda\Big(r^{n+1}+\frac{2}{3}(\frac{r^{n+4}}{n+4}-\frac{{r_i}^{n+4}}{n+4})\Big)\\\label{N48}
\end{eqnarray}

\begin{figure}
\center\epsfig{file=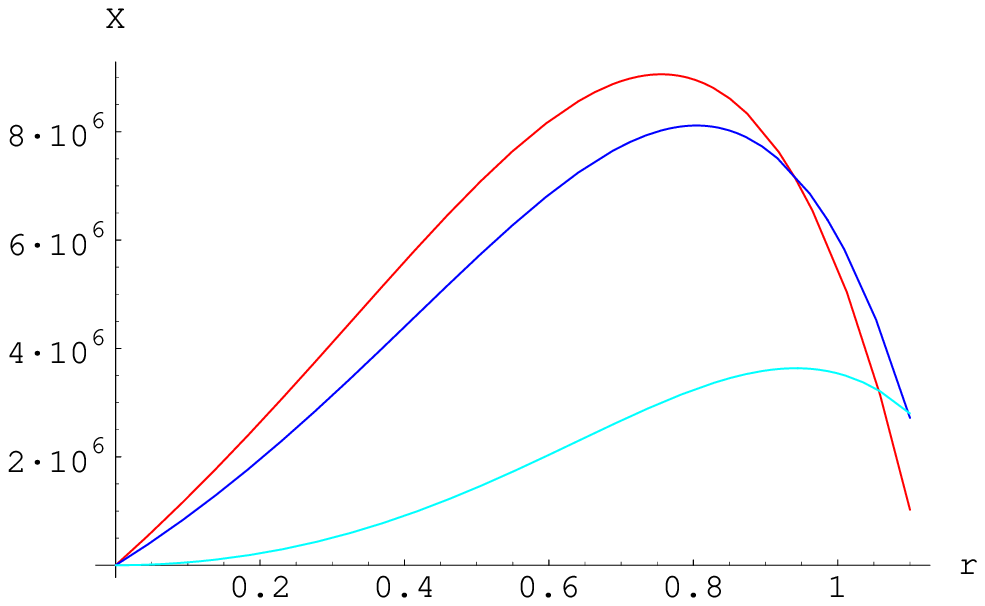, width=0.45\linewidth} \epsfig{file=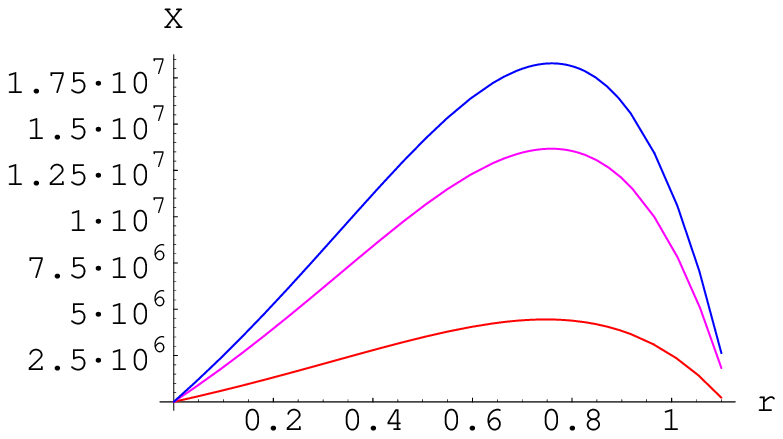,
width=0.45\linewidth}\caption{The left graph is plotted for
$\alpha=1,2,2.5$ and right graph is plotted for $c_0=-1,-3,-5$. For
both graphs $ m_0=10$, $(P_{r0}-P_{\perp 0})=5, P_{\perp 0}=10$ are
common} \center\epsfig{file=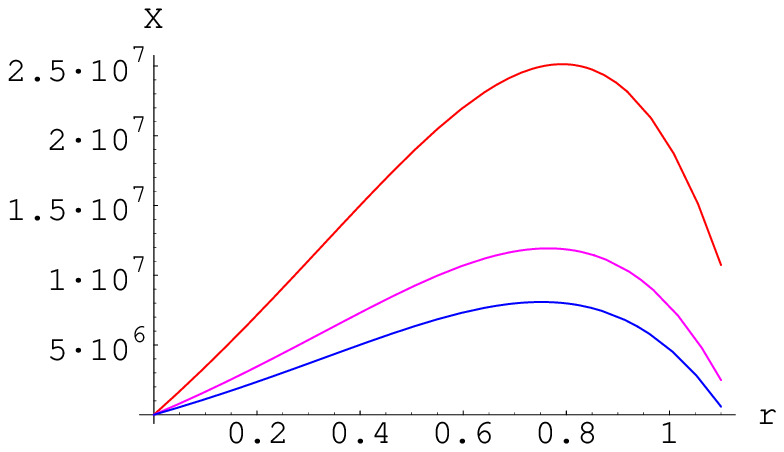,
width=0.45\linewidth}\epsfig{file=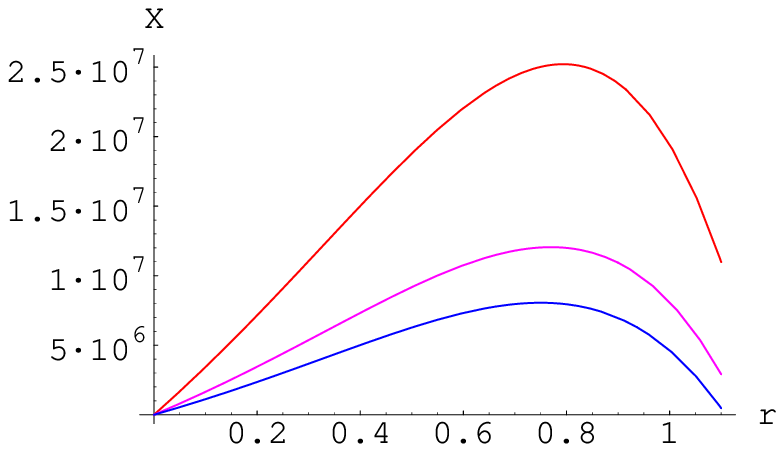,
width=0.45\linewidth}\caption{The left graph is plotted for
$m_0=9.5, 10.5, 12$ and right graph is plotted for $(P_{r0}-P_{\perp
0})=2,4,6$. For both graphs $ \alpha=1$, $c_0=-2, P_{\perp 0}=10$
are common.} \epsfig{file=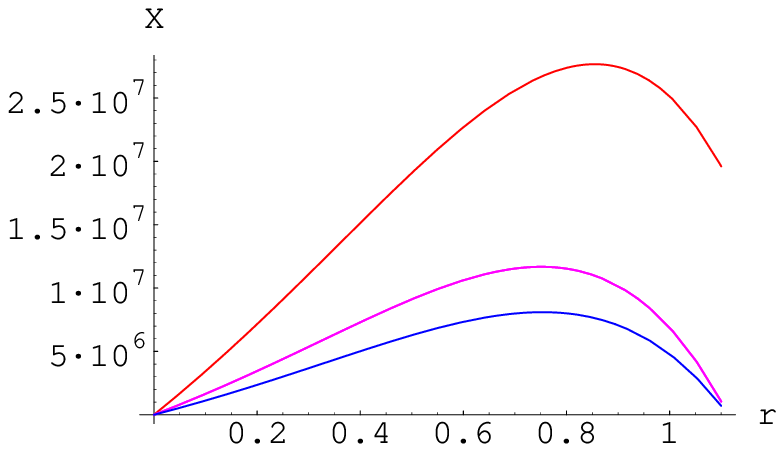, width=0.45\linewidth}\caption{This
graph is plotted for $(P_{r0}-P_{\perp 0})=2$, $ \alpha=1$, $c_0=-2,
P_{\perp 0}=10,12,13$.}
\end{figure}

\begin{figure}
\center\epsfig{file=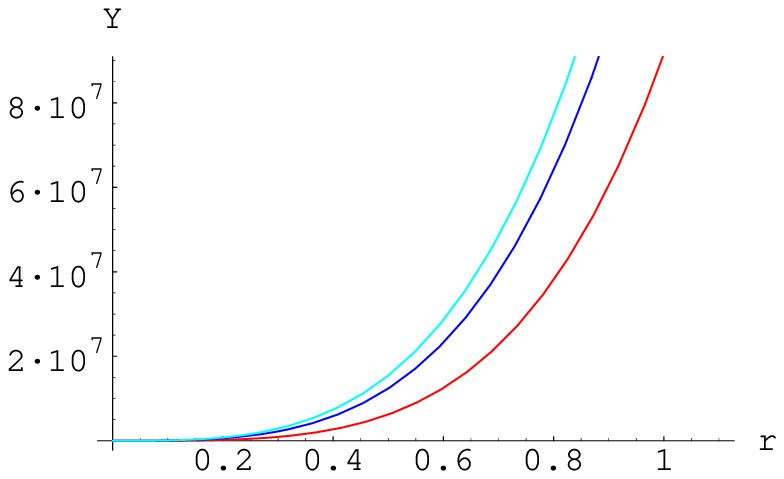, width=0.45\linewidth}
\epsfig{file=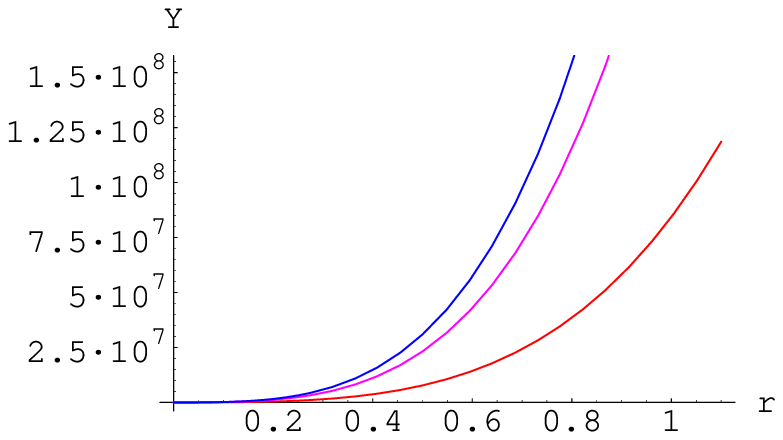, width=0.45\linewidth}\caption{The left graph is
plotted for $\alpha=1,2,2.5$ and right graph is plotted for
$c_0=-1,-3,-5$. For both graphs $ m_0=10$, $\lambda=2, n=4$ are
common} \center\epsfig{file=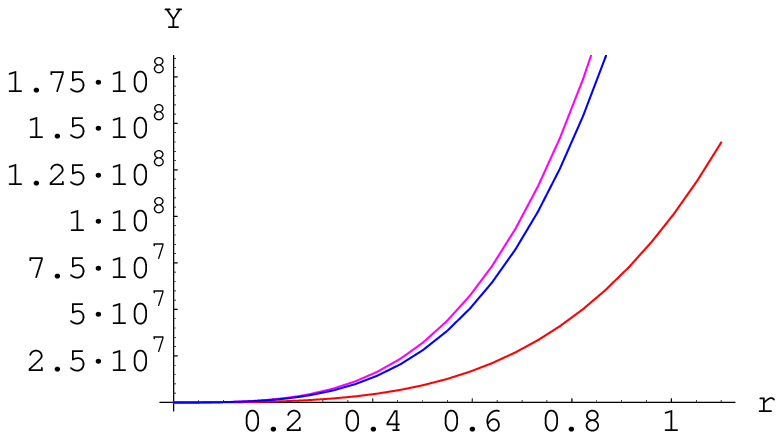, width=0.45\linewidth}
\epsfig{file=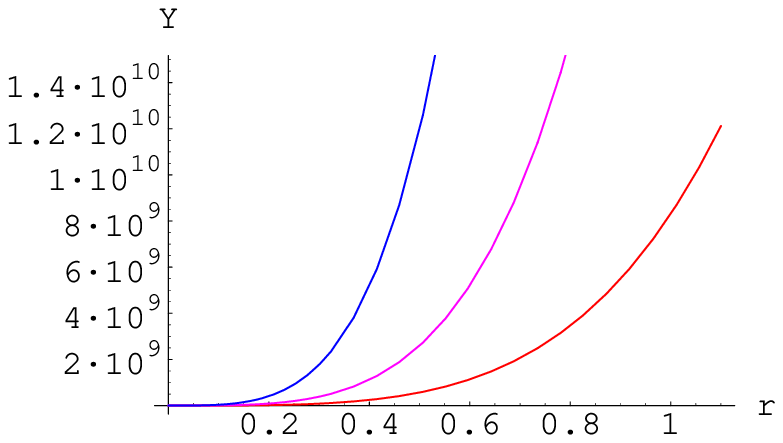, width=0.45\linewidth}\caption{The left graph is
plotted for $r_i=0.5, 0.7, 0.9$ and right graph is plotted for
$n=2,4,6$. For both graphs $ \alpha=1$, $c_0=-2, \lambda=2, m_0=10$
are common.} \center\epsfig{file=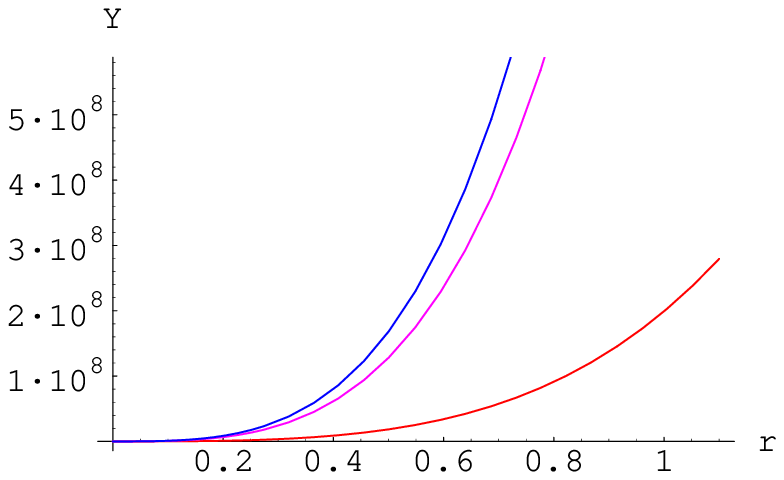, width=0.45\linewidth}
\caption{ This graph is plotted for$ \alpha=1$, $c_0=-2, n=2,
m_0=10, \lambda=2,4,6$ }
\end{figure}
Here, we have used Eq.(\ref{N16a}) and considered an energy density
profile of the form $\mu _{0}=\lambda r^{n},$ where $\lambda $ is
positive constant and $n$ is also a constant whose value ranges in
the interval $-\infty <n<\infty .$ In order to fulfill the stability
of expansion free fluids, we have to prove that both sides of
Eq.(\ref{N48}) produces positive results, which is analytically
impossible. We represent left side of Eq.(\ref{N48}) as $X(r)$ and
right side of this equation by $Y(r)$. We prove graphically that for
particulary values of the parameters involved in Eq.(\ref{N48}) both
$X(r)$ and $Y(r)$ positive. The positivity of $X(r)$ and $Y(r)$ is
shown in figures \textbf{(\textbf{1-3})} and
\textbf{(\textbf{4-6})}, respectively. The values of the parameters
for which $X(r)$ and $Y(r)$ remain positive (system predicts range
of stability) are mentioned below each graph and other than these
values system becomes unstable.
%
%
%
%
%
\section{Summary}
This paper deals with dynamical instability of the expansionfree
anisotropic fluid at Newtonian and post Newtonian order in the frame
work of Einstein Gauss-Bonnet gravity, which is vast play ground for
higher dimensional analysis of general relativity . For a
gravitating source which has non zero expansion scalar, the
instability range of a self gravitating source can be defined by the
adiabatic index $\Gamma_1$, which measures the compressibility of
the fluid under consideration. On the other hand, for an
expansionfree case as we are dealing, the instability explicitly
depends upon the energy density, radial pressure, local anisotropy
of pressure and Gauss-Bonnet coupling constant $\alpha$ at Newtonian
approximation, but it appears to be independent of the adiabatic
index $\Gamma_1$. In other words the stiffness of gravitating source
at Newtonian and post Newtonian approximation does not play any role
for the investigation of the stability of system. We would like to
mention that anisotropy in pressure, inhomogeneity in the energy
density and Gauss-Bonnet coupling constant $\alpha$ are the key
factors for studying the the structure formation as well as
evolution of shearfree anisotropic astrophysical objects.

We have formulated two dynamical equations how gravitating objects
evolve with time? and what is the final outcome of such evolution?.
One of these dynamical equations is used to separate the terms which
have Newtonian and post Newtonian order by using the concept of
relativistic and c.g.s units. The post post Newtonian regimes are
absent in the present analysis, it not due to Gauss-Bonnet gravity,
it seems to occur due to the geodesic properties of the spacetime
used in which $g_{00}=1$. This condition is in fact Newtonian limit
of general relativity. The second dynamical equation is used to
discuss the instability range of expansionfree fluid upto pN order.

The first order perturbation scheme has been applied on the metric
functions and matter variables appearing in the Gauss-Bonnet field
equations and dynamical equations. The analysis of resulting
dynamical equations shows that stability is independent of adiabatic
index $\Gamma_1$ due to expansionfree fluid. The instability depends
on the density profile, local anisotropy, Gauss-Bonnet coupling
constant and some other parameters. The instability required that
resultant of all term on left side of equation (\ref{N48}) should be
positive and equal to resultant of all terms on right side of that
equation. It is impossible to show analytically from Eq.(\ref{N48}),
so we have proved this result for a particular values of the
parameters appearing in Eq.(\ref{N48}). The domain of the parameters
is taken conveniently to show both sides positive Fig.
\textbf{(1-6)}. The parameters have following values for which
system satisfies stability conditions: $1\leqslant\alpha2.5$,
$-4\leqslant c_0\leqslant-1$,$9.5\leqslant m_0\leqslant12$,
$2\leqslant(P_{r0}-P_{\perp 0})\leqslant6$, $10\leqslant P_{\perp 0}
\leqslant13$, $0.5\leqslant r_i\leqslant0.9$, $2\leqslant\lambda15$,
$4\leqslant n\leqslant 8.$ We have the novel values of the
parameters one can carry actual calculations for the values of the
parameters by introducing some restrictions on the system under
consideration. This work with electromagnetic and heat flux in the
presence of non-geodesic model i.e., $g_{00}\neq1$ is under progress
\cite{58}.

\end{document}